\documentclass[aps,pra,twocolumn,amsmath,amssymb,superscriptaddress]{revtex4-1}
\usepackage{graphicx, subfigure}

\begin{document}

\author{Alan O. Jamison}
\email{jamisona@uw.edu}
\affiliation{University of Washington Department of Physics, Seattle, Washington 98195, USA}

\author{J. Nathan Kutz}
\affiliation{University of Washington Department of Physics, Seattle, Washington 98195, USA}
\affiliation{University of Washington Department of Applied Math, Seattle, Washington 98195, USA}

\author{Subhadeep Gupta}
\affiliation{University of Washington Department of Physics, Seattle, Washington 98195, USA}

\date{\today}

\begin{abstract}
We present theoretical tools for predicting and reducing the effects of atomic interactions in Bose-Einstein condensate (BEC) interferometry experiments. To address mean-field shifts during free propagation, we derive a robust scaling solution that reduces the three-dimensional Gross-Pitaevskii equation to a set of three simple differential equations valid for any interaction strength. To model the other common components of a BEC interferometer---condensate splitting, manipulation, and recombination---we generalize the slowly-varying envelope reduction, providing both analytic handles and dramatically improved simulations. Applying these tools to a BEC interferometer to measure the fine structure constant, $\alpha$ (Gupta, et al., 2002), we find agreement with the results of the original experiment and demonstrate that atomic interactions do not preclude measurement to better than part-per-billion accuracy, even for atomic species with relatively large scattering lengths. These tools help make BEC interferometry a viable choice for a broad class of precision measurements.
\end{abstract}

\title{Atomic Interactions in Precision Interferometry Using Bose-Einstein Condensates}
\maketitle

\section{Introduction}
\label{intro}
Interferometry with atomic beams or laser-cooled gases enables a number of high-precision measurements for applied and fundamental research \cite{cronin09}: rotation sensing \cite{durfee06}, measuring atomic properties \cite{miffre06}, and testing foundational principles of general relativity \cite{fray04} and quantum mechanics \cite{chapman95,*kokorowski01}, to name a few. Bose-Einstein condensates (BECs) have properties valuable to precision experiments: long coherence lengths (and correspondingly narrow momentum distributions) and high density. These ``atom lasers'' thus have tremendous potential for high-accuracy measurements.

Several promising experimental results have been shown with free-space \cite{gupta02,doring10,debs10} and confined (in traps or wave guides) \cite{wang05,veng07,burke08} BEC interferometers. While much theoretical work has been done to understand the effects of atomic interactions in confined interferometers \cite{olshanii05,stickney08,grond10,impens09}, the quality of the trapping potential tends to dominate considerations for high-accuracy measurements. In this work we present theoretical techniques to deal with atomic interactions in {\em free-space} BEC interferometers where these interactions are the fundamental hurdle for high-accuracy measurements.

In this paper we derive and apply two complementary techniques which together address the general problem of interaction effects: Robust scaling solutions correctly model BEC free expansion at any interaction strength, and a decomposition method efficiently models multi-path interferometers. Together these techniques allow modeling of a broad class of condensate interferometry experiments, both analytically and with simplified simulations. With these techniques it is now possible to explore a wide range of parameters in a short time to optimize experimental design as well as to run high-precision simulations of chosen parameters on commodity PCs. Our results provide key new tools for precision BEC interferometry.

We consider a large class of experiments in which a condensate is split (i.e., placed in a superposition of multiple momentum states), subjected to some differential phase shift, and then recombined. To keep the discussion focused, we will apply these techniques to a BEC interferometer to measure the fine structure constant, $\alpha$ \cite{gupta02}, which uses standing-wave laser diffraction pulses as mirrors and beam splitters for the condensate \cite{gupta01}. Our techniques agree both with full numerical simulations and the results of the first-generation contrast interferometry experiment.

For the class of experiments considered in this paper, collective excitations such as solitons, vortices, and shockwaves \cite{pethick08} are confined to the short time windows of condensate splitting and recombination and can be monitored with numerical simulations \cite{chang-engels08,*carretero-gonzalez08}. We find no such excitations in any of the simulations reported here. Further, densities and atom numbers are low enough to keep elastic collisions infrequent and below the threshold for bosonic stimulation \cite{band00}. This leaves the mean-field shift---the energy shift of an atom due to its interactions with all other atoms in the BEC---as the most important interaction effect for our work. While mean-field effects may, in principle, be reduced using Feshbach resonances\cite{gustavsson08,*fattori08}, doing so introduces systematic Zeeman shifts and the need to make nonzero magnetic fields ultra-stable and uniform. Further, the residual interaction effects need analysis, for which our techniques are valuable.

The paper is organized as follows. In section \ref{scaling_section} we discuss a robust analytic approximation for calculating the mean-field shift. In section \ref{SVEA_section} we develop a generalized slowly-varying envelope approximation. In section \ref{full_exps} we  apply our tools to identify experimentally accessible parameters such that mean-field effects do not preclude a sub-part-per-bllion (sub-ppb) measurement of $\omega_{\rm rec}$ and $\alpha$.

\section{Robust Scaling Solution to the Gross-Pitaevskii Equation}
\label{scaling_section}

In discussing atomic interaction effects, we will confine ourselves to the zero-temperature Gross-Pitaevskii description of a BEC, which is quite accurate for many interferometry experiments \cite{pethick08}. This approximation assumes that all atoms are in the condensate and that the many-body wavefunction has a form $\left.\psi(\vec{x_1},\ldots,\vec{x_N}t)=\prod_j \phi(\vec{x_j},t)\right.$. The Gross-Pitaevskii equation (GPE) is the mean-field approximation for $\phi$ and its temporal dynamics:
\begin{equation}
\label{GPE}
	i\hbar \frac{\partial \phi}{\partial t}= -\frac {\hbar^2}{2m} \nabla^2\phi + V(\vec{x})\phi + g|\phi|^2\phi ,
\end{equation}
where $g=4\pi\hbar^2a_s(N_{\rm at}-1)/m$, $a_s$ is the $s$-wave scattering length and $N_{\rm at}$ is the total atom number.

In this section we describe an approximate solution to the GPE which is accurate for both small and large interaction strength. Our scaling solution builds from a previous result valid in the Thomas-Fermi (TF) approximation, derived by Castin and Dum \cite{castin-dum96}. The TF approximation describes condensates where the interaction energy is much larger than the kinetic energy and may be considered as a semi-classical limit. Castin and Dum derived a ``dynamical TF'' approximation by explicitly considering this semi-classical nature, using an analogy to the expansion of a classical interacting gas to find a self-similar scaling solution to the GPE. Using an alternate derivation we find scaling solutions accurate near the origin for any interaction strength, which reduce to the result of Castin and Dum in the TF regime. In particular, our scaling solution reproduces the exact solutions both in the $g \rightarrow \infty$ (TF) limit and for $g=0$ (noninteracting).

For precision interferometry experiments, bringing the mean-field shift as low as possible without compromising the advantages of a BEC is essential, so the residual interaction effects will likely be well outside the regime where TF is applicable. As will be seen below, interaction effects for such BECs can still be large enough to spoil the results of precision interferometers.

We will consider the time evolution of a condensate in free space \footnote{The case of a condensate in a trap can be derived in the same fashion. Applying such solutions to interferometers in traps is the focus of ongoing work.}. Suppose the initial state has some given form $\phi_0\left( \vec{x}\right)$. Now, consider an ansatz of the form
\begin{equation}
	\label{ansatz}
	\phi\left( \vec{x},t \right) =
	\frac{\phi_0\left( \frac{x_1}{\lambda_1(t)},\frac{x_2}{\lambda_2(t)},\frac{x_3}{\lambda_3(t)} \right) }{\left(\lambda_1(t) \lambda_2(t) \lambda_3(t) \right)^{\frac 1 2} }
	e^{-i\theta\left(\vec{x},t\right)}\ ,
\end{equation}
where $\lambda_i$, $\theta$, and $\phi_0$ are all real valued. In the following, we will use the notation $y_i\equiv x_i/\lambda_i$ to simplify the appearance of equations. This gives $\partial / \partial x_i \rightarrow \lambda_i^{-1}\partial/\partial y_i$ and means there is some time dependence to $\phi_0(\vec{y})$ due to the $\lambda_i$'s. We substitute this ansatz into \eqref{GPE} and separate into two equations---one for the imaginary part and one for the real part. The equation for the imaginary part gives:
\begin{align}\label{ImPart}
	\sum_{j=1}^3\Bigg[& \frac {\phi_0} 2 \left( \frac{\dot\lambda_j}{\lambda_j}
	+\frac \hbar {m\lambda_j^2}\frac{\partial^2 \theta}{\partial y_j^2}\right) \notag \\
	&+ \frac{\partial \phi_0}{\partial y_j}\left(\frac{\dot\lambda_j y_j}{\lambda_j}
	+\frac \hbar {m\lambda_j^2}\frac{\partial \theta}{\partial y_j}\right)\Bigg] =0 \ .
\end{align}
Setting each of the two expressions in parentheses equal to zero we find the following condition on $\theta$:
\begin{equation}\label{theta_condition}
	\theta\left(\vec{y},t\right)=f(t)-\frac m {2\hbar} \sum_{j=1}^3 \dot\lambda_j\lambda_jy_j^2\ .
\end{equation}
With this form for $\theta$, the real part of the GPE yields
\begin{equation}
\label{RePart}
	\dot f-\frac m {2\hbar} \sum_{j=1}^3\lambda_j \ddot\lambda_j y_j^2
	= \frac{g\phi_0^2}{\hbar\lambda_1 \lambda_2 \lambda_3 }
	-\frac \hbar {2m} \sum_{j=1}^3\frac 1 {\lambda_j^2\phi_0}\frac{\partial^2 \phi_0}{\partial y_j^2}\ .
\end{equation}
Thus, if \eqref{RePart} can be satisfied exactly for a given $\phi_0$, the ansatz gives us an exact solution to the GPE.

We first look at the two extremes of interaction strength. For the TF limit of a condensate released from a harmonic trap at $t=0$, we have
\begin{equation}
\label{TF}
	\phi_0^2(\vec{y})=\frac 1 g \left(\mu-\frac 1 2 m\sum_{j=1}^3\Omega_j^2 y_j^2\right),
	\quad (g \rightarrow \infty)\ ,
\end{equation}
where $\Omega_j/(2\pi)$ is the trapping frequency in the $j$ direction and $\mu$ is the chemical potential in the trap. The second term on the righthand side of \eqref{RePart} becomes negligible, recovering the original result of Castin and Dum. In this limit, \eqref{RePart} can be solved exactly. As noted above, this result may be interpreted as the expansion of a classical, interacting gas driven by pressure due to repulsive interactions. From this perspective, the second term can be viewed as the first quantum correction, corresponding to an extra quantum ``pressure'' due to wavepacket dispersion.
\begin{figure*}
	\includegraphics[width=180mm]{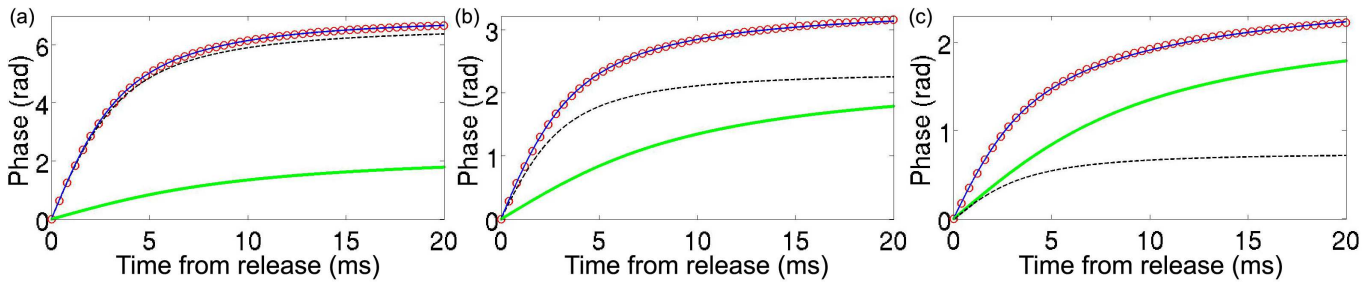}
	\caption{(Color online) Accuracy of scaling solutions at center of BEC. The phase at the center of the condensate is plotted versus expansion time for (a) high density, (b) intermediate density, and (c)low density. The red circles show results from full numerical simulations, while the solid blue curves show our scaling solutions using the initial condensate wavefunction calculated in the trap. For reference, the TF scaling solutions are plotted as dashed black curves and the noninteracting solution is plotted as thick green curves. Neither the TF nor the noninteracting solution is sufficient for high-accuracy measurements at any of these densities.}
	\label{tphase}
\end{figure*}
\begin{figure}
	\includegraphics[width=86mm]{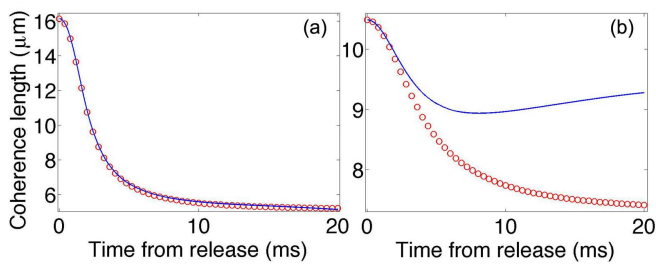}
	\caption{(Color online) Scaling solutions for coherence length. The coherence length is plotted versus expansion time for (a) high density and (b) intermediate density. The red circles show results from full numerical simulations, while the solid blue curves show our scaling solutions using the initial condensate wavefunction calculated in the trap. The coherence length is not well approximated with the scaling solutions for intermediate densities where the ${\cal O}(y^2)$ expansion breaks down.}
	\label{coh_length}
\end{figure}

In the opposite limit of a noninteracting gas (or a single atom) released from a harmonic trap, $\phi_0$ is gaussian and again \eqref{RePart} can be solved exactly. Thus, we can expect \eqref{ansatz} subject to \eqref{theta_condition} and \eqref{RePart} to give good results for very small or very large $g$. To extend this solution to all $g$ values, we view the above equations as an expansion in $y$. Thus we find solutions valid for all interaction strengths in a region near the center of the condensate.

The signal in most interferometry experiments will be dominated by the highest density region when the arms overlap. This is true because the majority of atoms will be found here. Also, with tiny offsets from perfect overlap, as are inevitable in experiments, the rapid phase oscillations in the wings will largely cancel any contributions far from the slowly-varying peak density region. Furthermore, since the primary goal is to understand how to suppress the phase shift due to mean-field effects, setting an upper bound is key. Accurately knowing the mean-field phase shift at the center of the condensate, where it will be greatest, sets a tight upper bound.

Therefore, an expansion valid for small deviations from the origin should capture most of the relevant physics. We expand the righthand side of \eqref{RePart} in a power series in $y$ and keep only terms up to order $y^2$. Equating the coefficients for the $y^0$ terms and the $y^2$ terms yields the following pair of equations for $f$ and the $\lambda$'s:
\begin{equation}
	\label{feq}
	\dot f =\frac g \hbar \frac {\alpha_0} { \lambda_1 \lambda_2 \lambda_3 }
	-\frac {\hbar} {2m} \sum_{k=1}^3\frac {\alpha_k} {\lambda_k^2 }
\end{equation}
\begin{equation*}
	\alpha_0\equiv \left(\phi_0(0) \right)^2 \quad , \quad
	\alpha_k\equiv\frac 1 {\phi_0(0)}\left.\frac{\partial^2 \phi_0}{\partial y_k^2}\right|_{\vec{y}=0}
\end{equation*}
\begin{equation}
	\label{lambdaeq}
	\ddot\lambda_j= \frac {g} {m}\frac {-\beta_{0j}} { \lambda_j\lambda_1 \lambda_2 \lambda_3}
	+\frac {\hbar^2} {2m^2}\frac 1{\lambda_j}
	\sum_{k=1}^3\frac {\beta_{kj}} {\lambda_k^2}
\end{equation}
\begin{equation*}
	\beta_{0j}\equiv\left.\frac{\partial^2\phi_0^2}{\partial y_j^2}\right|_{\vec{y}=0} \quad , \quad
	\beta_{kj}\equiv\left.\frac{\partial^2}{\partial y_j^2}
	\left[\frac 1 {\phi_0}\frac{\partial^2\phi_0}{ \partial y_k^2} \right]\right|_{\vec{y}=0}
\end{equation*}
where the $\alpha$'s and $\beta$'s are constants calculated from the initial state.

To demonstrate the utility of these equations we compare full simulations of the three-dimensional GPE to approximate solutions obtained from the above equations \footnote{In full simulations all time evolution was performed in
  momentum space using fourth-order, adaptive Runge-Kutta and fast Fourier
  transforms. Initial in-trap states were found by imaginary-time evolution to
  find the lowest energy steady state. All numerical results were performed with
  multiple grid sizes to check convergence of numerical solutions. Grid sizes
  were increased until the differences between results of simulations run on
  different grids were negligible on the scale of the results presented.}.
We simulated the evolution of a $^{174}\rm{Yb}$ ($a_s = 5.6\;{\rm nm}$) BEC released from a harmonic trap with frequencies $\left(\Omega_x,\Omega_y,\Omega_z\right) = 2\pi\times \left(50,50,20\right)\;{\rm Hz}$. The condensate was allowed to expand for $20\;{\rm ms}$. This is the relevant time-scale for the interferometer to measure $h/m$ at the sub-ppb level described in section \ref{full_exps}.

Figure \ref{tphase} shows the phase at the center of the condensate as a function of time for condensates with $N_{\rm at} = 10^4$, $10^3$, and $10^2$ (corresponding to high, intermediate, and low interactions). In the early stages of the expansion with $N_{\rm at} =10^4$, the TF result agrees well with the full numerical solution. This shows the initial density (peak density in trap of $6.2\times10^{13}\;{\rm cm^{-3}}$) falls within the TF regime. However, we see that the TF result begins to diverge from the numerical solution after around $5 \rm ms$, growing to a substantial deviation at the end of the expansion. This is an important result for precision measurements. While a condensate may begin an experiment in the TF regime, its subsequent expansion lowers the density, eventually making TF no longer a good approximation. For condensates even deeper in the TF regime (such as the scenario discussed in section \ref{full_exps}) this feature becomes even more pronounced due to the faster expansion.

For $N_{\rm at} = 10^2$ (peak density in trap of $6.5\times10^{12}\;{\rm cm^{-3}}$) the low density makes the TF approximation inaccurate throughout. However, the scaling solutions derived above maintain good validity. A common suggestion for precision BEC interferometry is to adiabatically lower the in-trap density before beginning an experiment. The deviation of the $N_{\rm at} = 10^2$ results from both TF and the noninteracting case highlights the need to be able to analyze interaction effects even for seemingly low initial densities.

The middle case of $N_{\rm at} = 10^3$ (peak density in trap of $2.4\times10^{13}\;{\rm cm^{-3}}$) shows marked departure from both the TF and noninteracting results. The continued agreement with our scaling solution in the intermediate regime highlights the robustness of this technique.

To further discuss the extent of validity of our scaling solutions, we now address more global properties. Density and phase profile across the condensate generally show good agreement for all three interaction regimes. These properties individually are not usually of direct experimental importance for atom interferometry. Instead, we present here the more relevant parameter of coherence length $l_{\rm c}$ which depends sensitively on both density and phase profiles. For a condensate wavefunction $\phi\left(\vec{x}\right)$ we define the coherence length:
\begin{equation}
\frac{\int d^3x\; \phi\left(\vec{x}\right)^*\phi\left(\vec{x} \pm (l_{\rm c}/2)\hat{x_3}\right)}{\int d^3x\; \phi\left(\vec{x}\right)^*\phi\left(\vec{x}\right)} = \frac 1 e\ .
\end{equation}
For a two-arm interferometer this is directly related to the $1/e$ coherence time of the signal through the relative velocity of the two arms at recombination \cite{hagley99}. For a three-arm interferometer there is a similar relationship, up to some numerical factors related to the splitting.

As figure \ref{coh_length} shows, the scaling solutions reproduce this global property well for high density ($N_{\rm at} = 10^4$) but show clear ($25\%$ in fig \ref{coh_length}b) deviations for intermediate densities ($N_{\rm at} = 10^3$). These deviations grow to $50\%$ for the $N_{\rm at} = 10^2$ case and then decrease sharply to below $5\%$ for $N_{\rm at} = 10^1$. That our scaling solutions are inadequate to describe the coherence length at intermediate interaction strengths is not surprising given that this is a global quantity, and our scaling solutions are only accurate near the origin. In the small interaction case, the deviations fall in line with the relative size of the $y^2$ and $y^4$ terms in the small $y$ expansion of the right-hand side of \eqref{RePart} found using first-order perturbation theory. This suggests that the failure to accurately reproduce the coherence length in intermediate densities directly follows from truncating the expansion at ${\cal O}(y^2)$. We will show in section \ref{full_exps} how our method accurately reproduces phase offsets for even the difficult case of a long experiment that moves from deep in the TF regime to weak interactions. Thus, extension of the scaling technique beyond ${\cal O}(y^2)$ is beyond the scope of this work.


\section{Slowing-Varying Envelope Approximation}
\label{SVEA_section}

The class of BEC interferometers we consider typically contains widely disparate scales. Laser pulses used to manipulate condensates have durations from hundreds of nanoseconds to tens of microseconds \cite{gupta01}, while the entire experiment can last for tens or hundreds of milliseconds. Accurate simulation with such a separation of scales is computationally intensive---in some cases it may be prohibitively intensive. We decompose the condensate wavefunction into small, quasi-independent pieces, making simulations dramatically less computationally expensive. Our decomposition technique, known as the slowly-varying envelope approximation (SVEA), is well-known in the context of optics \footnote{The use closest in
	application to the present case comes in modeling wavelength division multiplexing systems. See \cite{agarwal06}.}. Its application to BECs was pioneered by Trippenbach et al \cite{trippenbach00}.

The SVEA leverages one of the key experimental advantages of a BEC: its narrow momentum spread. The momentum-space wavefunction of a condensate typically has a width well below the recoil momentum of a constituent atom due to absorption of a visible photon. Thus, when the condensate wavefunction is split using a light grating, the momentum-space wavefunction consists of a series of clearly separated peaks. The SVEA may be applied to any splitting method that creates clearly separated momentum states, and thus is applicable to many possible experiments.

We generalize the SVEA, showing that certain terms dropped in the usual description of the technique can be important at the level of accuracy needed for precision experiment modeling. We also identify a consistent expansion whose small parameter gives us an estimate of the accuracy of this method.

To start, we postulate a form for the condensate wavefunction:
\begin{equation}
	\label{SVEA ansatz}
	\phi\left( \vec{x},t\right) =\sum_{j}\phi_j\left( \vec{x}-\frac {\hbar\vec{k}_j}{m}t,t
	\right)e^{i\vec{k}_j\cdot\vec{x}-i\omega_j t}
\end{equation}
where the $\vec{k}_j$'s are the relevant wavevectors and the $\omega_j$'s the corresponding frequencies. We will consider $\left.\vec{k_j}=2j\vec{k}_{\rm rec}\right.$ where $j$ is an integer and $\vec{k}_{\rm rec}$ is the laser wavevector. After inserting this ansatz we reorganize the GPE into the following suggestive form:
\begin{widetext}
\begin{equation}
\label{SVEAGPE}
	\sum_{j=-\infty}^\infty i\hbar\frac{\partial \phi_j}{\partial t}
	e^{i\vec{k}_j\cdot\vec{x}-i\omega_j t}
	= \sum_{j=-\infty}^\infty\left[-\frac{\hbar^2}{2m}\vec{\nabla}_{\xi_j}^2\phi_j
	+g\!\!\!\!\!\sum_{l_1,l_2,l_3=-\infty}^\infty \!\!\!\!\!
	\phi^*_{l_1}\phi_{l_2}\phi_{l_3}\delta_{j,-l_1+l_2+l_3} e^{-i\left(-\omega_j-\omega_{l_1}
	+\omega_{l_2}+\omega_{l_3}\right)t} \right]
	e^{i\vec{k}_j\cdot\vec{x}-i\omega_j t} \quad,
\end{equation}
\end{widetext}
where coordinates $\vec{\xi}_j=\vec{x}-\hbar\vec{k}_jt/m$ were chosen for the $\phi$'s and $\delta_{a,b}$ is the Kronecker delta function. The choice of coordinates cancels a term of form $\vec{k}_j\cdot\vec{\nabla}\phi_j$, and the usual non-relativistic relation $\omega=\hbar k^2/(2m)$ is used to cancel another pair of terms.

Consider the Fourier transform of this equation. If all $\phi_j$'s (``envelopes'') have bounded support centered at zero with diameters smaller than one third
	\footnote{For a linear equation one expects the condition to be one rather than one third, which is equivalent to the $\phi_j \exp\left(i\vec{k}_j\cdot \vec{x}\right)$ terms being orthogonal. The stronger condition is necessary to keep momentum separation in the nonlinear term. This can be extended to higher nonlinearities, with a fifth order nonlinearity necessitating the stronger ``one fifth'' condition on the support of the $\phi_j$'s and so on.}
	of the minimum of $|k_i-k_j|$, then this equation separates exactly into an infinite tower of equations of the form
\begin{align}
\label{phi_j_eq}
	i\hbar \frac{\partial \phi_j}{\partial t}
	= &-\frac{\hbar^2}{2m}\vec{\nabla}_{\xi_j}^2\phi_j \notag \\
	&+g\!\!\!\!\!\!\sum_{l_2+l_3-l_1=j} \!\!\!\!\!\!
	\phi^*_{l_1}\phi_{l_2}\phi_{l_3}e^{-i\left(\omega_{l_2}+\omega_{l_3}-\omega_{l_1}-\omega_j\right)t} \ .
\end{align}
This equation differs slightly from that found in the literature \cite{agarwal06}. Derivations of the SVEA refer to collecting ``phase matched'' terms into separate equations. In the context of optics where $\omega_j \propto k_j$, spatial phase matching is equivalent to temporal phase matching. For matter waves with $\omega_j \propto k_j^2$ the two conditions may be considered separately. We have retained pieces with $\left. \vec{k}_{l_2}+\vec{k}_{l_3}-\vec{k}_{l_1}-\vec{k}_j=0 \right.$ but $\left. \omega_{l_2}+\omega_{l_3}-\omega_{l_1}-\omega_j\neq 0 \right.$ (i.e., we enforce spatial but not temporal phase matching). Such pieces can have large enough effects to be important in simulating precision experiments. Also, they are essential for including laser interactions in the SVEA.

Given initial conditions satisfying the above criterion, this tower of equations is equivalent to the full GPE. To simplify we select a subset of the envelopes to keep and set the rest to zero. This truncation affects the envelopes we do keep by dropping terms from the nonlinear piece. We want to know how large an error this induces.

Since we only set $\phi_m \equiv 0$ if $\phi_m(t=0)=0$, for short times we can drop the kinetic term to find
\begin{equation*}
	i\frac{\partial \phi_m}{\partial t}
	= \frac g \hbar \!\!\!\!\sum_{l_2+l_3-l_1=m} \!\!\!\!\!\!
	\phi^*_{l_1}\phi_{l_2}\phi_{l_3}e^{-i4\omega_{\rm rec}t \left(l_2^2+l_3^2-l_1^2\right)} \ .
\end{equation*}
For time scales short enough that the other envelopes do not appreciably move or expand, this can be integrated. The initially unpopulated terms would oscillate with frequency at least $4\omega_{\rm rec}$ and amplitude smaller than the initially populated states by the ratio of the mean-field energy to the recoil energy. This result gives us a good error estimate, since envelopes expanding and moving relative to one another will decrease the righthand side, and also suggests a way to systematically improve the accuracy of the approximation by retaining one or more of the initially unpopulated terms.

\begin{figure}
	\includegraphics[width=85mm]{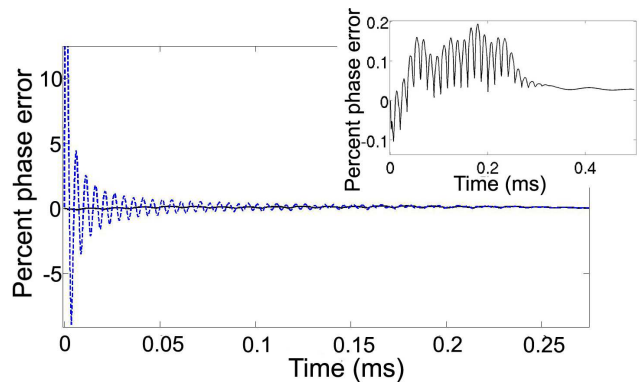}
	\caption{(Color online) Errors in SVEA from temporal phase matching. The phase at the center of the zero-momentum branch of the condensate in SVEA simulations is compared to the result of full simulations of the GPE. The blue dashed curve shows the SVEA with both spatial and temporal phase matching. The black solid curve shows the modified SVEA with only spatial phase matching. The initial peak of the dashed curve rises to 26\%, but has been cropped from the image to make other details visible. The inset shows the solid curve on a scale where details are visible.}
	\label{SVEA_compare}
\end{figure}
To verify these uncertainties, we ran simulations using a condensate wavefunction with an initial superposition of three momentum states ($0$ and $\pm 2 \hbar k_{\rm rec}$), all populated with equal density. For these simulations we used Na condensates ($a_s=2.9\;{\rm nm}$) with $N_{\rm at} = 10^4$ and trap frequencies $\left(\Omega_x,\Omega_y,\Omega_z\right) = 2\pi\times \left(50,50,20\right)\;{\rm Hz}$. The results from SVEA simulations were compared to results from full simulations of the GPE. Figure \ref{SVEA_compare} shows the fractional error in the phase accumulated by the zero momentum branch during separation. The percent error seen in simulations with both spatial and temporal phase matching fluctuates with a frequency $2\times(4\omega_{\rm rec})$ just as would be expected for error due to neglecting a term $\left.\phi_0^*\phi_1\phi_{-1}e^{-i(4\omega_{\rm rec})t(1^2+1^2)}\right.$. The drop-off of fluctuations just after 0.2 ms corresponds to complete separation of the condensate branches.

External potentials may be folded into the SVEA approach. The potentials created by laser standing waves used for splitting and acceleration deserve special mention. The light-shift potential formalism shows that standing-wave laser pulses may be described by a potential of the form
\begin{equation}
\label{laser_potential}
	V(\vec x,t)=A(\vec x,t)\left|e^{i\vec k_{\rm rec}\cdot \vec x}+e^{-i\vec k_{\rm rec}\cdot \vec x} \right|^2
\end{equation}
where $A$ is the amplitude of the potential, which may depend on time and space. The use of relative detunings may create potentials moving relative to the lab frame. In the rest frame of such a potential, the $e^{\pm 2i\vec k_{\rm rec}\cdot \vec x}$ pieces connect branches of the condensate wavefunction with momenta differing by $\pm 2\hbar \vec k_{\rm rec}$. Since dynamics during a laser pulse occur at the time-scale of the recoil frequency, removing the $e^{-i\omega_j t}$ phase is no longer acceptable. Then \eqref{phi_j_eq} is modified by the reappearance of the term $(\hbar k_{\rm rec})^2\phi_j/(2m)$, which was canceled by the time dependence of $e^{-i\omega_j t}$ in the original derivation. Thus, for modeling laser interactions it is key that temporal phase-matching {\em not} be enforced \footnote{ Accurate simulation of the physics during laser
     interactions requires keeping track of a number of initially unpopulated condensate branches. We have found that for both Kapitza-Dirac and Bragg pulses it is sufficient to consider two extra accessible branches on each side of the range you expect to populate (for better than percent-level accuracy of all final wavefunctions). For instance, a Bragg pulse that takes $2\hbar k_{\rm rec}$ to $-2 \hbar k_{\rm rec}$ will also require keeping track of the $0$ momentum branch, two more states above $2\hbar k_{\rm rec}$ and two more states below $-2 \hbar k_{\rm rec}$. However, once a laser interaction is complete, the branches that are no longer populated can be removed from the simulation, keeping the number of states tracked from growing during simulation of an experiment with many light gratings.}.

\section{Modeling a Complete Experiment}
\label{full_exps}
In this section we use our theoretical tools to model a contrast interferometer to measure $\omega_{\rm rec}$ and $\alpha$. The first-generation of this experiment was reported in \cite{gupta02}. This scheme makes a useful test case since it strongly leverages the unique advantages of BECs for precision measurement.
\subsection{Contrast Interferometry Scheme for $\omega_{\rm rec}$ and $\alpha$}
\begin{figure}
	\includegraphics[width=85mm]{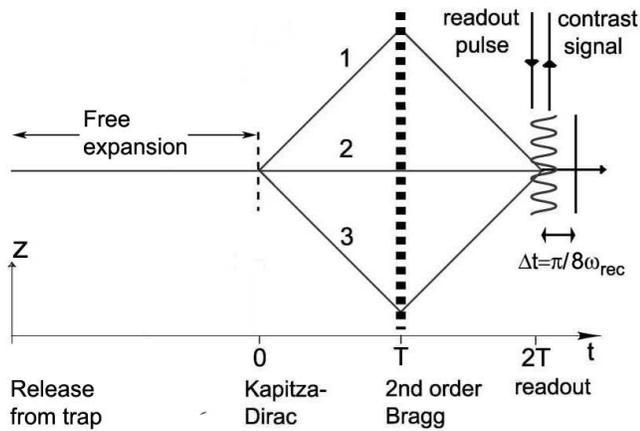}
	\caption{Scheme of the contrast interferometer for $\omega_{\rm rec}$ and $\alpha$. Using diffraction gratings of light, the condensate is split into three branches of momenta $2\hbar k_{\rm rec}$, $0$, and $-2\hbar k_{\rm rec}$ and then recombined after a variable time $2T$. When all three overlap, they create a matter-wave grating whose contrast rises and falls with frequency $8\omega_{\rm rec}$. }
	\label{oldCI}
\end{figure}
In this interferometry scheme (see figure \ref{oldCI}), a condensate is split into three branches with momenta $2\hbar k_{\rm rec}$, $0$, and $-2\hbar k_{\rm rec}$ by a short-duration standing-wave light grating (a ``Kapitza-Dirac pulse'') \cite{gupta01}. The three branches evolve phase according to their kinetic energies. After a time $T$, a long-duration light grating (``Bragg pulse'') is applied to the condensate, causing the non-zero momentum states to reverse direction. At time $2T$ all three states again overlap to create a matter-wave interference pattern with contrast that oscillates with time. This time-varying contrast is measured using a traveling-wave readout laser pulse with the same wavevector as the light gratings, which in simulations corresponds to the $2\hbar k_{\rm rec}$ component of the density.

The readout laser pulse is Bragg reflected from the matter-wave grating. The interferometer signal is the intensity of this reflection, which scales with the contrast of the matter-wave grating and is proportionate to $\left.\sin^2\left((\phi_1+\phi_3)/2 -\phi_2 \right)=\sin^2\left(4\omega_{\rm rec}t +\phi_{\rm offset}\right)\right.$, where $\phi_i$ is the phase of branch $i$ (see figure \ref{oldCI}), $\left.\omega_{\rm rec}=\hbar k_{\rm rec}^2/(2m)\right.$ is the recoil frequency, and $\phi_{\rm offset}$ is time-independent and does not affect the final measurement. We define $\phi_{\rm sig}\equiv (\phi_1+\phi_3)/2 -\phi_2 $. The value of $\omega_{\rm rec}$ is extracted from the variation of $\phi_{\rm sig}$ with $T$. Finally, the recoil frequency may be used to determine the fine structure constant $\alpha$ \cite{bouchendira11}.

The first-generation experiment---which used a Na BEC, $T = 3\;{\rm ms}$, and momenta $\pm 2 \hbar k_{\rm rec}$ in branches 1 and 3---reported $7$ ppm precision, but an inaccuracy at the $200$ ppm level \cite{gupta02}. Simple estimates indicated that this large inaccuracy likely arose from the differential mean field shift between the three arms of the interferometer. Since finite densities are needed for adequate signal-to-noise and since the arm imbalance cannot be arbitrarily controlled, residual mean-field interactions will always be present. This has
been a strong motivation for the current work which is aimed at a thorough quantification
of the mean field effect for a ppb-level measurement of the fine structure constant. We specifically consider use of Yb BECs to extend this technique to sub-ppb precision \cite{jamison11}. Using additional Bragg pulses to give momenta of $\pm 40 \hbar k_{\rm rec}$ to branches 1 and 3 and $T=5\;{\rm ms}$, the total phase accumulation $\phi_{\rm sig}\approx3.7\times10^5\;{\rm rad}$. With $N_{\rm at}=10^4$ the shot-noise limit is $0.01\;{\rm rad}$. This allows accumulation to sub-ppb precision in less than a day. Other systematic errors to this measurement can be kept within the sub-ppb range with reasonable choices of parameters for the diffracting laser beams \cite{jamison11}.
Within these parametric constraints, we find that the mean-field shift may be reduced to below one billionth of $3.7\times10^5\;{\rm rad}$, such that atomic interactions do not preclude a sub-ppb measurement of $\omega_{\rm rec}$.
\subsection{Simulations of Complete Interferometers}
Two full experiments were simulated to confirm the accuracy of our various techniques, after which we simulated the contrast interferometer of Gupta, et al \cite{gupta02}. The first allows us to test the validity of the SVEA decomposition as compared to full numerical solutions to the GPE, showing that the potentially dramatic reduction of computational cost allowed by the SVEA does not diminish the accuracy of simulations. The second shows the efficacy of our scaling solutions in several parameter regimes as compared to three-dimensional SVEA simulations, for experiments that could not be adequately simulated on commodity PCs using the full GPE.

\begin{figure}
	\includegraphics[width=85mm,height=44mm]{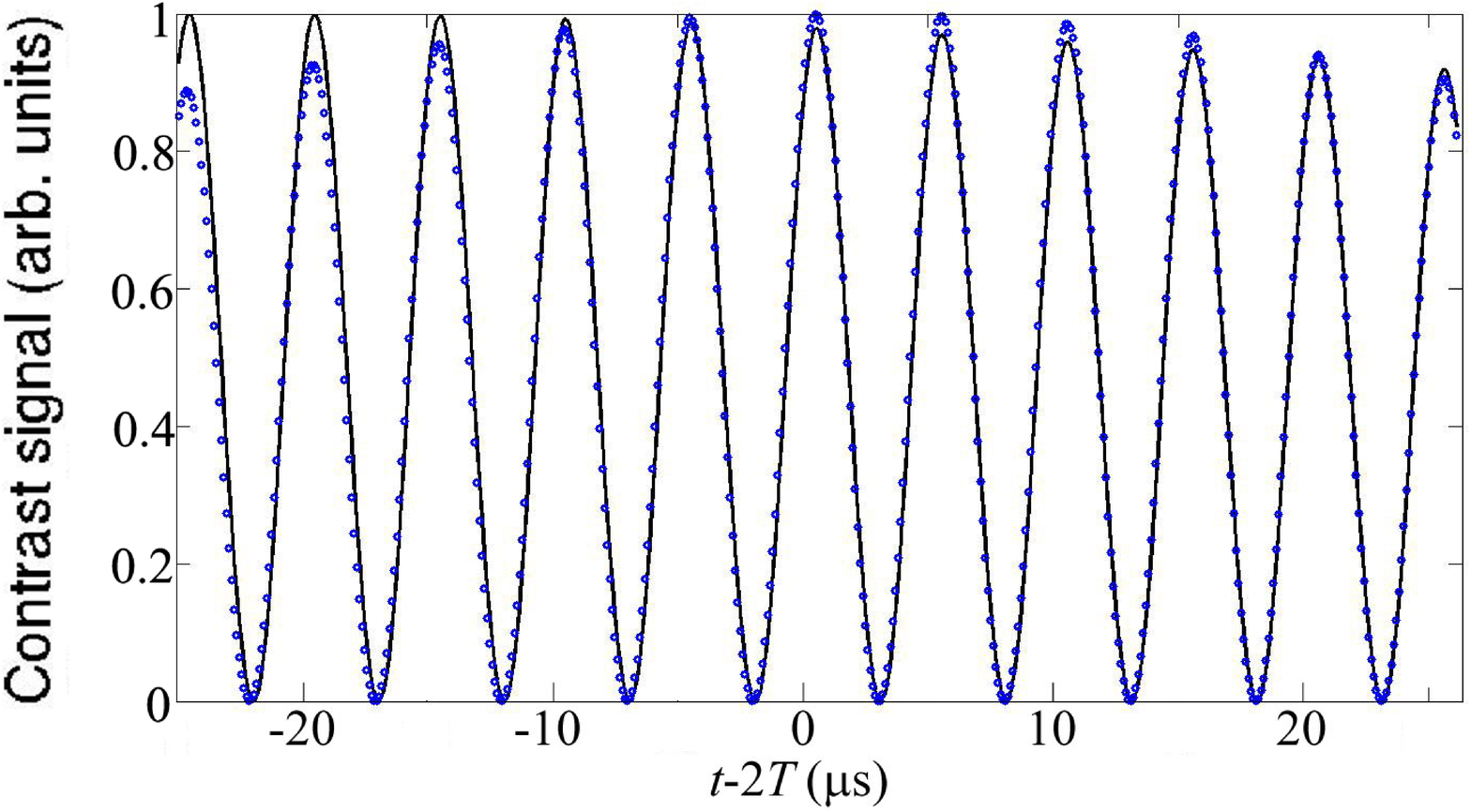}
	\caption{(Color online) Simulated output of a contrast interferometry experiment. The solid (black) curve shows the output signal simulated using the full GPE. The circles (blue) show the output signal simulated using the SVEA.}
	\label{GPEvsSVEA}
\end{figure}

First, we simulated a short experiment using a Na condensate with $1\;{\rm ms}$ of free expansion out of the trap ($\left(\Omega_x,\Omega_y,\Omega_z\right) = 2\pi\times \left(50,50,20\right)\;{\rm Hz}$) and $T=0.2\;{\rm ms}$. This experiment is sufficiently short that full simulations of the GPE may be run in reasonable time. Figure \ref{GPEvsSVEA} shows the signal from the full GPE simulation along with the signal from the SVEA simulation. This signal was generated by extracting the magnitude of the $2\hbar k_{\rm rec}$ component of the total atomic density as a function of time. The phase at time $2T$ agrees within the granularity of the time steps in the simulated signal \footnote{We believe the small differences in signal envelope arise from spurious gratings created by higher momentum states, which we remove from the SVEA simulations.}.
Since the final phase of this signal is sensitive to slight differences between simulations in the free propagation, condensate separation/recombination, or laser interaction periods, the agreement seen in figure \ref{GPEvsSVEA} shows the power of the SVEA to accurately model all periods of an experiment. The actual experiment will consider the slope of a $\phi_{\rm sig}$ vs. $T$ plot in which many of these details are expected to cancel out. Being able to accurately model these details can thus improve the confidence in these cancelations, allow modeling of laser intensity fluctuation effects, etc.
\begin{figure}
	\includegraphics[width=85mm]{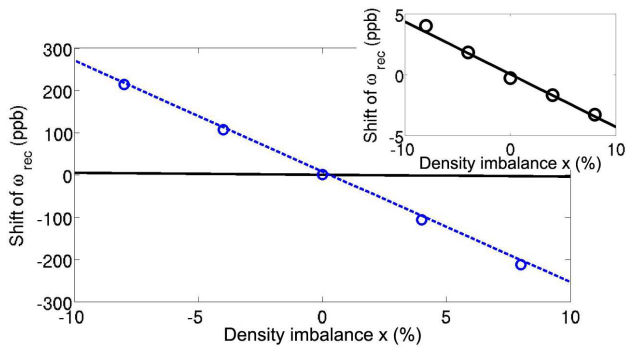}
	\caption{(Color online) Mean-field shifts for various splitting parameters. The black solid line shows the scaling solution value for the shift in measured recoil frequency as a function of density imbalance for the experiment described in the text with $\Omega = 2\pi\times200\;{\rm Hz}$. The blue dashed line shows the same but for an initial trapping frequency of $\Omega = 2\pi\times5\;{\rm Hz}$. The circles show data points from SVEA simulations. The inset shows a closer view of the solid line. For imbalance levels of $\leq 2\%$ the mean-field shift is reduced to the 1ppb level.}
	\label{omega_shift}
\end{figure}

Several recent studies of atom-light interactions with varying levels of complexity \cite{stamperkurn01,*ketterle01,buchner03,*muller08} can easily be adapted to describe the condensate splitting. In our framework, the complexity of the laser interaction model may be changed as needed without substantially affecting the models for the rest of the interferometer. For the simulations described above we used the light-shift potential formalism described in section \ref{SVEA_section}. For the following, we use a much simpler model ignoring phases accumulated during the laser interactions, as these should not affect the final result of an experiment.

The second simulated experiment is the proposed Yb contrast interferometer to measure $\alpha$ to sub-ppb precision described above. We assume $N_{\rm at}=10^4$ and the condensate initially confined in an isotropic trap with frequency $\Omega = 2\pi\times 200\;{\rm Hz}$. After a $10\;{\rm ms}$ period of free expansion, the BEC is split into three branches. The densities in branches 1, 2, and 3 are in the ratio $(1-x):(1+2x):(1-x)$, where $x$ is the difference in fraction of original density between the stationary branch and the accelerated branches (e.g., density splitting in the ratio 1:2:1 has $x=0.25$). We make use of our scaling solutions from section \ref{scaling_section} to quickly probe a large region of parameter space to find suitable conditions for the experiment.

To apply our scaling solutions, we break a full experiment into stages of free propagation and stages of laser interaction. While in section \ref{scaling_section} we only considered releasing the condensate from a harmonic trap at $t=0$, the scaling solutions may be used for more general situations by carefully choosing initial conditions. For expansion from a trap, the initial conditions are $f(0)=0$, $\lambda_j(0)=1$, and $\dot\lambda_j(0)=0$. After a Kapitza-Dirac pulse, the condensate wavefunction has the form
\begin{equation}
	\alpha \phi_{\rm KD}+\beta\left(\exp(2i\vec{k}_{\rm rec}\!\cdot\vec{x})
	+\exp(-2i\vec{k}_{\rm rec}\!\cdot\vec{x})\right)\phi_{\rm KD} \ ,
\end{equation}
where $\phi_{\rm KD}$ is the wavefunction immediately before the Kapitza-Dirac pulse \footnote{We have ignored the branches with higher magnitude momentum
  because SVEA simulations show that they contribute only small effects for the splitting parameters we have considered}.
If we ignore the interbranch terms from the SVEA, then the scaling equations may be used to continue evolving each branch of the wavefunction forward, using $\phi_0=|\phi_{\rm KD}|$, $f(0)=\arg\left(\phi_{\rm KD}(\vec{x}=0)\right)$, $\lambda_j(0)=1$, and $\dot\lambda_j(0)=(2\hbar/m) \partial^2[\arg(\phi_{\rm KD})]/\partial x_j^2|_{\vec{x}=0}$, as initial conditions.

To account for the interbranch interactions, the simplest approach is to treat each branch as though it is acted upon by a weak potential due to each other branch, ignoring the back-reaction---the change in shape of each branch due to its interactions with other branches. This approximation gives the first term of an expansion in a small parameter, which we describe for the case of release from a trap followed by some period of expansion before the splitting laser pulse, treating the initial expansion with the scaling solutions.

If the $z$ axis is the direction of laser propagation, we find the relative size of the neglected, second-order back-reaction effects to be $[(\Omega_z/\omega_{\rm rec}) (\mu/(\hbar \Omega_z) )^3]^{1/2}\lambda_x^{-1}\lambda_y^{-1}$ for a condensate in the TF regime and $[(\Omega_z/\omega_{\rm rec})(\mu/(\hbar \Omega_z) )^2]^{1/2}\lambda_x^{-1}\lambda_y^{-1}$ in the small interaction limit. With this approximation we treat the entire post-Kapitza-Dirac pulse propagation with a single scaling solution and add the interbranch interaction phase at the end. These parameters are generally small for BEC experiments (e.g., the first-generation contrast interferometry experiment \cite{gupta02} had $\Omega_z/ \omega_{\rm rec}<10^{-3}$) showing this to be a good approximation.

We find the fractional shift of the measured recoil frequency due to atomic interactions as a function of $x$ (see figure \ref{omega_shift}) by calculating the phase of the signal at $t=2T$ for runs of $T=2\;{\rm ms}$ and $T=5\;{\rm ms}$ and then finding the slope of $\phi_{\rm sig}$ vs. $T$. This agrees well with the results of SVEA simulations. For $|x|<0.02$ the mean-field shift contributes at less than the ppb level. Such precise control of the density splitting has been demonstrated by Hughes et al \cite{hughes07}.

For a given atom number and available total time for a run, the mean-field shift is generally smaller for larger trap frequencies because the condensate expands much more rapidly after release from strong traps than weak traps, quickly making up for the higher initial density. However, as trap frequency is increased, the final momentum spread after expansion increases, potentially diminishing the advantage of a BEC's narrow momentum distribution. At $\Omega = 2\pi\times 200\;{\rm Hz}$, the final momentum spread is still less than one tenth of the recoil momentum, while allowing us to bring the mean-field shift comfortably below the ppb level.

Finally, with these confirmations in hand, we simulated the 2002 experiment of Gupta, et al \cite{gupta02}, in which Na condensates of $N_{\rm at}\sim 10^5$ were split with $x\approx 0.25$. Using our scaling solutions we find an expected relative shift due to mean-field of $-3\times10^{-4}$, which compares well with the $-2\times10^{-4}$ systematic error reported by Gupta, et al.

\section{Conclusions}
\label{conclusions}
The use of a BEC source has the potential to improve precision atom
interferometry measurements for fundamental and applied research, once atomic
interaction effects are tamed. While a general rule to achieve this is to simply work with
reduced density, doing so may compromise other aspects of the experiment such as
interferometer signal-to-noise. We have presented new theoretical techniques to precisely
quantify interaction effects allowing their analysis in high accuracy measurements at the
sub-ppb level using BEC interferometers.

We derived scaling solutions to the Gross-Pitaevskii equation valid for all
interaction strengths in the central region of the condensate---the part which dominates
most atom interferometry experiments. We demonstrated their validity by comparing
against numerical simulations of the 3D GPE.

Using our modified slowly-varying envelope approximation, we have shown
the ability to rapidly simulate BEC interferometry experiments that utilize standing-wave gratings and beam splitters. We demonstrated negligible errors in the final read-out signal of the contrast interferometer for $\alpha$ when compared against numerical simulations of the 3D GPE. Our scaling solutions adapted with correction terms from our SVEA analysis allow for precise estimation of the contrast interferometer phase. Large regions of parameter space (as demonstrated
for the arm imbalance parameter) can be quickly analyzed to identify suitable conditions for high accuracy measurements, which would have prohibitive computational expense using full simulations of the GPE. The mean field shift predicted by this method is consistent with the magnitude and sign of error reported in the first-generation contrast interferometer experiment [6].

We have shown that atomic interaction effects are not a road-block to a sub-ppb measurement of $\omega_{\rm rec}$ and $\alpha$ with condensates. The presented methods can be useful in precision BEC interferometry applications where mean-field effects need to be well understood and the consequent measurement uncertainty properly accounted for.

\begin{acknowledgements}
This work was supported by the National Science Foundation under grants DMS-1007621 and PHY-0906494 and by a National Institute of Standards and Technology Precision Measurement Grant. We thank J. Vinson for helpful discussions and D.M. Stamper-Kurn for helpful comments on the manuscript.
\end{acknowledgements}

\end{document}